\documentclass[a4paper,11pt]{article}
\pdfoutput=1 

\usepackage{jinstpub} 

\title{A study of radiation tolerance in optical cements}

\author[1]{R.J.~Tesarek,\note{Corresponding author.}}
\author{E.~Hahn,}
\author{A.~Pla-Dalmau,}
\author[2]{J.L.~Salinas~Jr.\note{Present Address: University of Illinois, Chicago, 
Chicago, IL 60607}}

\affiliation{Fermi National Accelerator Laboratory, Batavia, IL 60119, USA}
\emailAdd{tesarek@fnal.gov}

\abstract{We study the effect ionizing radiation has on light transmission 
in the wavelength range 190--1100~nm for a number of optically clear epoxies. 
We find that the transmittance of traditional, commercially available, optical 
epoxies show significant degradation for exposures of 
$1\times10^{12}$~MIPs/cm$^2$.  Degradation of light transmission progresses 
from the shortest wavelengths at low doses to longer wavelengths as the dose 
increases.  In epoxy joints that are 0.1~mm thick, we 
observe that more than 5\% of the light is lost for wavelengths less than
400~nm for traditional optical epoxies. Our studies have identified 
an optically clear epoxy that shows little degradation for radiation exposures 
up to $5.9\times 10^{14}$~MIPs/cm$^2$ ($\approx 220$~kGy).}


\arxivnumber{326189} 
\notoc

\begin{document}
{\flushright
FERMILAB-PUB-20-276-E \\
}
\maketitle
\section{Introduction}
\label{sec:intro}
Plastic scintillators and light sensors have a long tradition for measuring 
radiation in particle and nuclear physics. Speed, linearity, ability to cover 
large areas with low channel count and ability to machine into complex shapes 
have made this technology a good choice for large detector 
systems~\cite{ScintDetRef}. Current and proposed experiments that run at
high radiation exposure rates and over long experiment lifetimes mean that 
detector systems must survive ever more hostile radiation environments.  
Considerable information is available in the literature that characterizes 
radiation tolerance of scintillator materials~\cite{Zorn1993, CMS-HCAL2016} 
and photosensors~\cite{PMTdamage,SiPMdamage,APDdamage}.  However, 
much less information is available on the radiation tolerance of optical
coupling compounds that connect scintillators and their 
sensors~\cite{Buechel2017, Kirn1999, Huang1997, Para1991}.  Our 
observations from actual devices operating in high radiation environments 
are that the adhesive in the joints are visually darker than either the plastic
scintillator or acrylic light guide material.  In this paper we explore
the radiation tolerance of a variety of commercially available, optically
clear epoxies.  We wish to identify candidate clear epoxies with improved
radiation tolerance for light transmission in the wavelength region 190--1100~nm.

The ideal optical epoxy would have good adhesion and bond strength, transmit 
light in the optical region, similar refractive index as the materials
used to make the detectors, and a reasonable open time to allow for assembling
the detectors before the adhesive agent hardens.  For this initial study, we
chose a few epoxies with the following properties:
\begin{itemize}
 \itemsep0em
 \item Readily available from commercial vendors.
 \item The epoxy is advertised as clear.
 \item Bond strength of at least 100 kg/cm$^2$.
 \item Working time greater than 10 minutes.
 \item If known, a refractive index of approximately 1.5.
\end{itemize}
For comparison we chose two commercially available optical epoxies designed for
use with plastic scintillator materials, BC-600 epoxy from 
Saint-Gobain~\cite{BC-600} and EJ-500 from Eljen Technologies~\cite{EJ-500}.  
We purchased two epoxies from System~3~\cite{System3}, a general purpose 
marine epoxy\footnote{The general purpose epoxy has three choices for hardener,
15, 30, and 60 minute working times for hardeners 1,2 and 3 respectively.  We 
use the medium cure time hardener (2) for the tests reported in this article.}
and a clear coating epoxy "Mirror Coat".  
Finally, we purchased two epoxies from Hysol~\cite{Hysol}, E-30CL and U-09FL.  

While most manufacturers offer considerable technical information on their
epoxies, one parameter often missing for optical applications is the
refractive index (R.I.).  The refractive index of the epoxy being of
critical importance to minimize Fresnel losses at the interfaces between
differing optical media. We measure the refractive index of all cured epoxy
samples using a Vee Gee model C10 Abbe 
refractometer~\cite{refractometer}\footnote{We note 
that the monobromonapthalene liquid typically used to optically couple the sample
to the glass in the refractometer is a solvent for many plastics and 
epoxies.}.
All refractive indices were measured at a single wavelength of
589~nm.  Table ~\ref{epoxyproperties} summarizes physical
properties of epoxies included in the studies reported here.  For more
details on the epoxies we refer to the technical data sheets available
from the manufacturers.
\begin{table}[tb]
 \begin{center}
 \caption{\label{epoxyproperties}Summary of epoxy properties from the 
  manufacturer's technical data sheets.  Some adhesive properties were 
  not available (N/A) from the manufacturer.  Values denoted by an asterix
  represent quantities measured rather than found in the appropriate technical
  data sheet. The refractive index reported for all epoxies is for a single 
  wavelength of 589~nm.}
  \begin{tabular}{lllrrrrl}
   &&&
   \multicolumn{1}{c}{Work} &
   \multicolumn{1}{c}{Hard.} &
   \multicolumn{1}{c}{Mixed} &
   \multicolumn{1}{c}{Bond} &
   \multicolumn{1}{c}{} \\

   &&&
   \multicolumn{1}{c}{Time} &
   \multicolumn{1}{c}{Time} &
   \multicolumn{1}{c}{Visc.} &
   \multicolumn{1}{c}{Stren.} &
   \multicolumn{1}{c}{Spec.} \\

   \multicolumn{1}{c}{Name} &
   \multicolumn{1}{c}{Abbr.} &
   \multicolumn{1}{c}{R.I.}       & 
   \multicolumn{1}{c}{(min)} &
   \multicolumn{1}{c}{(hrs)} &
   \multicolumn{1}{c}{(cps)} &
   \multicolumn{1}{c}{($\rm kg/cm^2$)} &
   \multicolumn{1}{c}{Grav.} \\ \hline
    BC-600~\cite{BC-600}
      & BC      & 1.571 & 180 & 24 &  800 & 125 & 1.18 \\
    EJ-500~\cite{EJ-500}
      & EJ      & 1.574 &  60 & 24 &  800 & 125 & 1.17 \\
    S3:MirrorCoat~\cite{System3}
      & S3MC    & 1.561* &  40 & 72 &  700 & N/A & 1.14* \\ 
    S3:General~\cite{System3}
      & S3GP    & 1.567* &  30 &  6 & 1100 & 527 & 1.10 \\
    H:E-30CL ~\cite{Hysol}                                                    
      & HECL    & 1.520* &  30 & 2.6 & N/A  & 560 & 1.07 \\
    H:U-09FL~\cite{Hysol}
      & HUFL    & 1.49*  &  10 & 3 -- 24 &  N/A  & 194 & 1.0 -- 1.2 \\
   \hline  
  \end{tabular}
 \end{center}
\end{table}

The different epoxies were tested for compatibility with a standard 
poly(vinyltoluene) scintillator (Eljen EJ-200). The new epoxies were prepared and
poured into one of two 1.91~cm diameter forms attached to a 
$6.45\times 6.45\;{\rm cm^2}$ piece 
of scintillator.  The other form was filled with EJ-500 epoxy.  
Once the epoxies were cured, the sample was stored in a cabinet for future 
examination. After over a year had elapsed, the scintillator pieces were
examined under a microscope and the two sample areas compared 
specifically by looking for changes in the scintillator at the
epoxy/scintillator interface, eg. discoloration, micro-cracks (crazing).  For 
all epoxies studied here, we were unable to observe any degradation 
of the scintillator epoxy interface.

\section{Sample Preparation}
\label{sec:SamplePreparation}
Optical cement samples were carefully prepared in two batches separated
by 1 year using the prescription described below.  To ensure uniform sample 
sizes and geometries, forms were made by cutting 19~mm (3/4 inch) ID clear 
schedule 40 PVC pipe into 9.53~mm long sections.  The cut edges of each form 
were sanded smooth and cleaned with ethanol.  A thin mylar sheet was then
taped to a glass plate to provide a flat surface to which the forms would
be affixed.  The forms were then glued to the mylar using a cyanoacrylate
adhesive and allowed to cure while the optical epoxies were prepared.  All 
handling of any material contacting a sample during preparation or handling
of a sample as described below was performed with gloves and tweezers 
to avoid contamination.

Sufficient optical epoxy was mixed to make 8--12 samples, approximately 
50~ml total. All optical cements, epoxies were mixed according to their 
manufacturer's instructions. To remove air bubbles, each mixture was placed 
in a centrifuge and run at 3,200 rpm for 5 minutes. After the centrifuge 
the clear adhesive was poured into the forms so that a meniscus protruded 
above the top of each form.  The epoxy samples were allowed to cure for 
several days. 

Once the samples had cured, the forms were pealed from the mylar and 
each epoxy sample removed from its form.  A label was then written on the
cylindrical portion of each sample in indelible ink.  An orientation line
was also filed into the cylindrical edge. Each sample was then cut and
diamond polished on both sides to approximately 7~mm thick. We note that 
the samples are thick compared with typical glue joints of approximately 
0.1~mm thickness.  We will address the thickness issue in a later section.   
The HUFL epoxy has a cured consistency similar to silicone rubber.  Without 
substantial additional handling and possible contamination, the HUFL samples 
could not be diamond polished and therefore are thicker than the samples 
for other epoxies. The thickness of each sample
was measured in multiple places and an average thickness determined.
Fractional variation of sample thicknesses for any given epoxy ranged between
0.5\% to 6\%.  The average fractional thickness variation between different 
epoxies is 10\% if the HUFL samples are included and 0.5\% if the HUFL 
samples are excluded.  All prepared samples were then cleaned with ethanol 
and the label re-applied as needed.  Table~\ref{ThicknessTable} 
summarizes the sample thicknesses.

\begin{table}
 \caption{\label{ThicknessTable}Table of sample thickness measurements.  
  Samples were prepared in two batches with ID less than 20 for the first batch 
  and ID  larger than 20 for the second batch. The fractional variation in 
  thickness for each candidate epoxy was less than 6\% for all samples.}
 \centering
 \begin{tabular}{l|rrrrrr|}
  & \multicolumn{6}{c|}{Sample thickness (mm)} \\ \hline
  \multicolumn{1}{c|}{ID} &
  \multicolumn{1}{c}{BC} &
  \multicolumn{1}{c}{EJ} &
  \multicolumn{1}{c}{S3MC} &
  \multicolumn{1}{c}{S3GP} &
  \multicolumn{1}{c}{HECL} &
  \multicolumn{1}{c|}{HUFL} \\ \hline
        1 & 7.57 & 7.57 & 7.47 & 7.49 & --      & --    \\
        2 & 7.57 & 7.65 & 7.62 & 7.57 & --      & --    \\
        3 & 7.62 & 7.54 & 7.54 & 7.59 & --      & --    \\
        4 & 7.57 & 7.32 & 7.54 & 7.54 & --      & --    \\
        5 & 7.57 & 7.57 & 7.57 & 7.09 & --      & --    \\
        6 & 7.59 & 7.19 & 7.59 & 7.24 & --      & --    \\
      21 & 7.62 & 7.62 & 7.67 & 7.67 & 7.62 & 10.01 \\
      22 & 7.65 & 7.65 & 7.61 & 7.53 & 7.38 &   8.81 \\
      23 & 7.66 & 7.61 & 7.04 & 7.62 & 7.67 & 10.03 \\
      24 & 7.66 & 7.61 & 7.61 & 7.67 & 7.62 &   9.40 \\
      25 & --      & --     & 7.62 & --      & --     & --    \\
      26 & --      & --     & 7.62 & --      & --     & --    \\ \hline
    Avg & 7.61 & 7.53 & 7.54 & 7.50 & 7.57 & 9.56 \\ 
 Stdev & 0.04 & 0.15 & 0.17 & 0.19 & 0.13 & 0.58 \\\hline
 \end{tabular}
\end{table}

\begin{figure}[tb]
 \centering
 \includegraphics[width=0.45\textwidth]{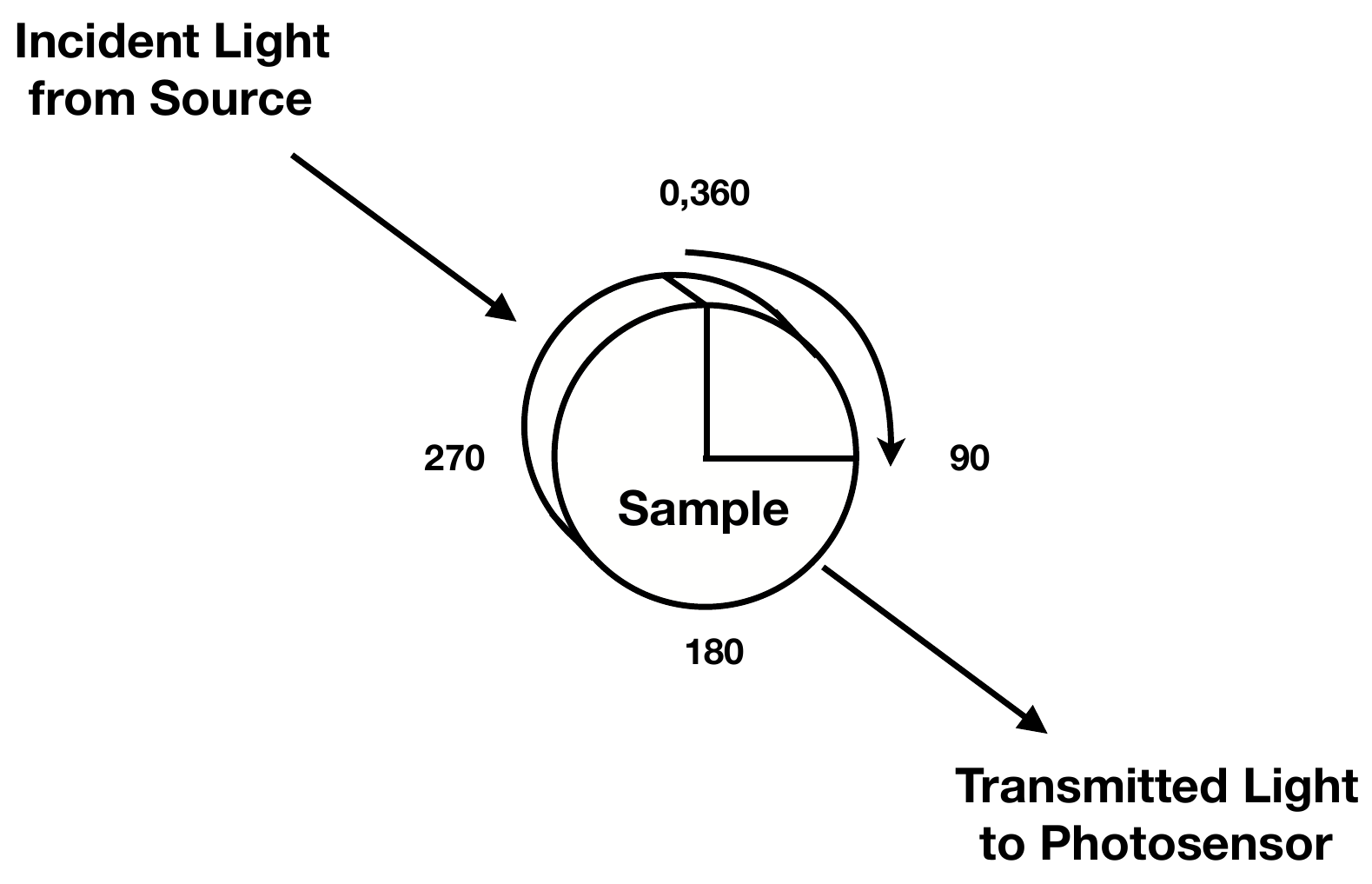}
 \caption{\label{SampleRotation} Schematic of sample orientations in the 
 spectrophotometer and representative transmittance curves for a single,
 unirradiated sample.}
\end{figure}

\begin{figure}[tb]
 \centering
 \includegraphics[width=0.45\textwidth]{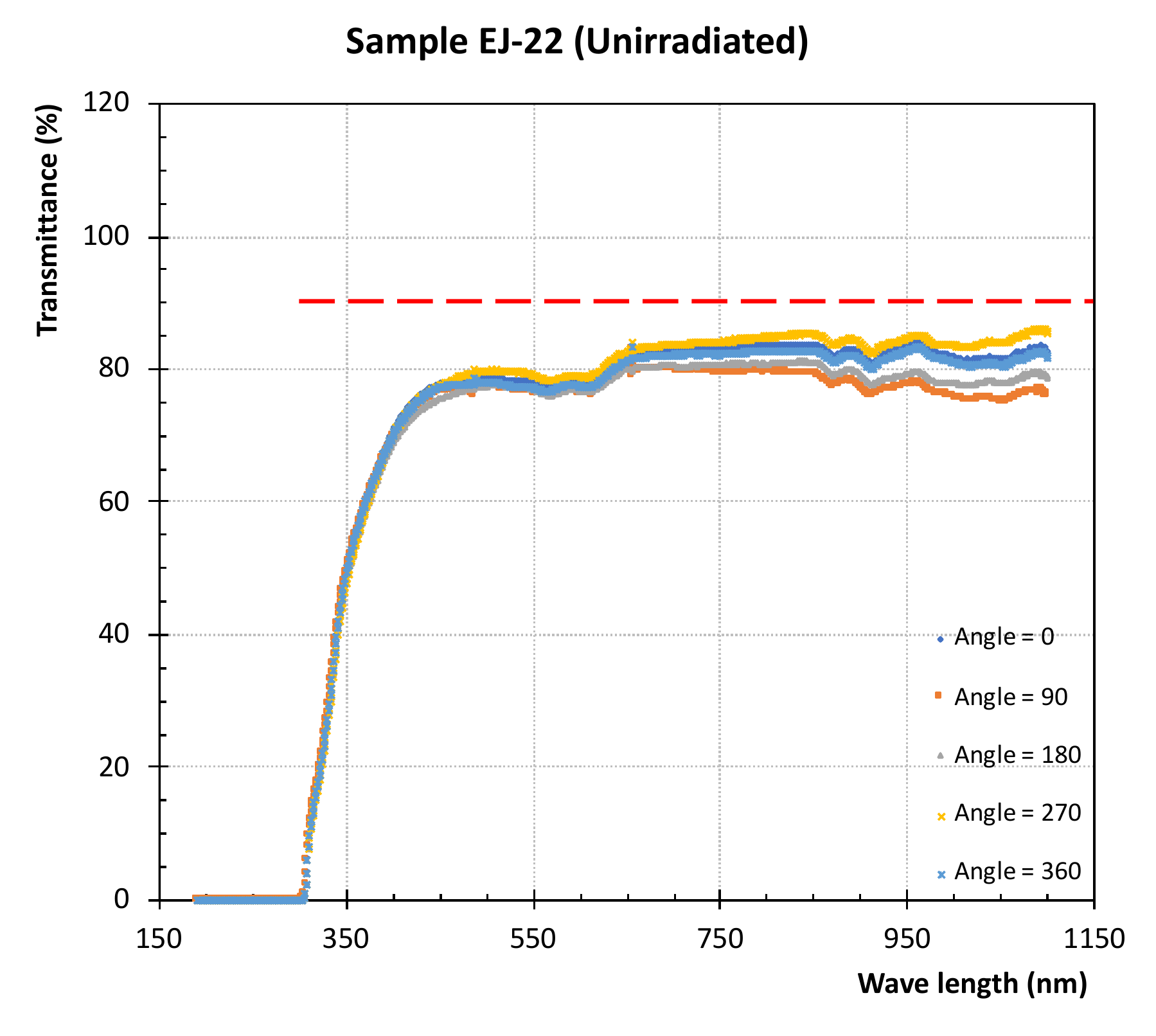}
 \includegraphics[width=0.45\textwidth]{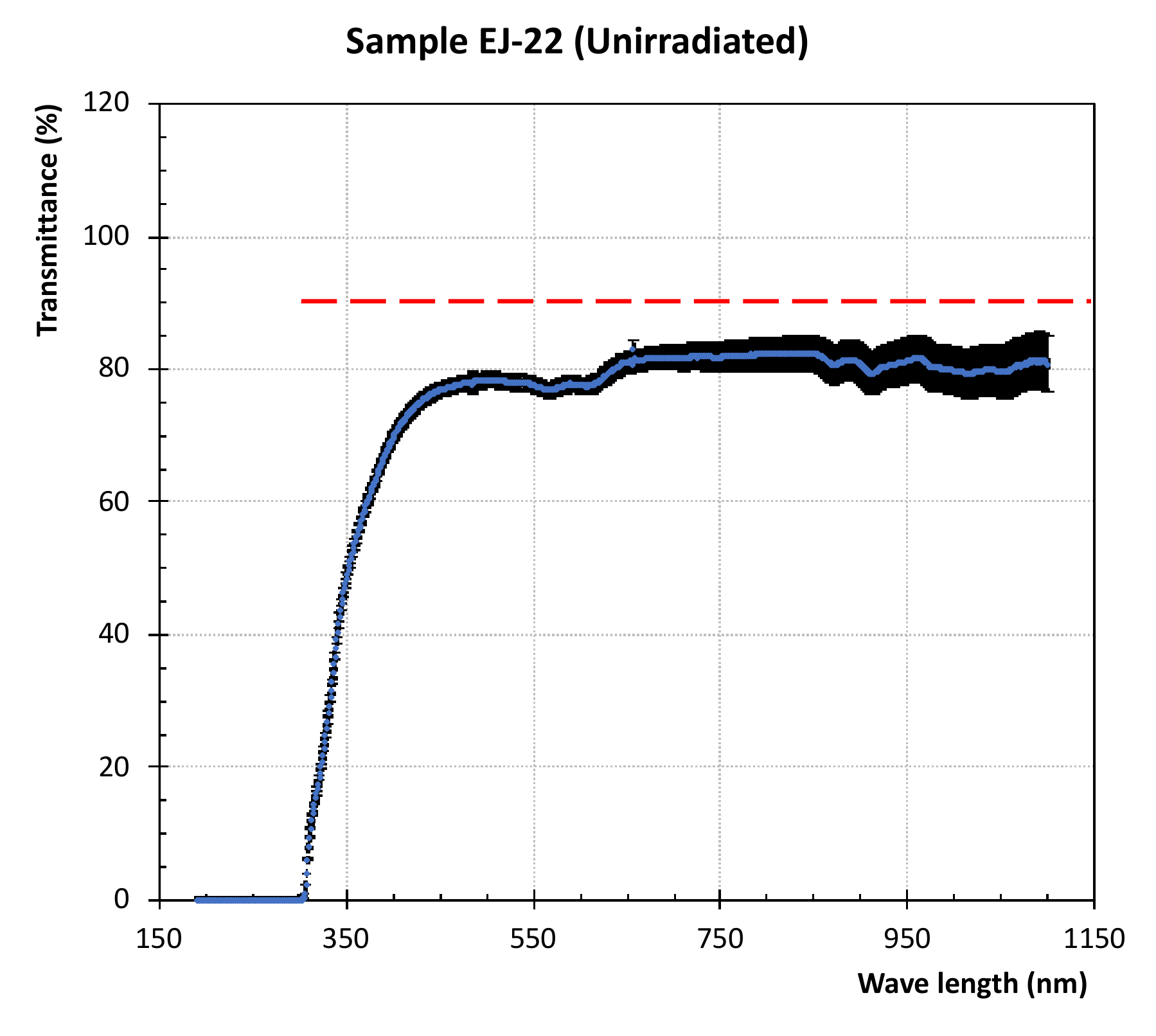}
 \caption{\label{SampleTransmittance}Raw transmittance data for 
 an unirradiated, EJ-500 sample showing all curves(left) and the average
 for rotations 0-270 degrees (right).  Note that the curves for rotations 0,360 
 degrees are nearly identical indicating good reproducibility of a given
 measurement.  The narrow feature at 650~nm is present in all transmittance
 measurements.  The dashed horizontal lines indicate the theoretical
 transmittance expected from Freznel losses only given the refractive index
 measured at 589~nm. }
\end{figure}

After preparation, the transmittance spectrum of each sample was measured
using an Agilent Technologies spectrophotometer~\cite{Spectrophotometer}.
Transmittance is defined as ratio of light intensity passing through the 
material to the incident intensity. The transmittance measurements covered 
the wavelength range 190~nm--1100~nm in 1~nm steps using a 0.5~s 
integration time for the optical sensor. The data were converted to a comma 
separated value format for subsequent analysis.  These spectra give an 
initial evaluation of the transmittance spectrum for each epoxy before 
irradiation.  Transmittance measurements were made of each sample
in 5 different orientations, each 90 degrees from the others. By doing so
we sample the same orientation twice to check consistency of the measurements.
The transmittance results of the different orientations are averaged to 
give a single value for each wavelength and the RMS of the measurements
were calculated. Figure~\ref{SampleRotation} shows the sample 
orientations for the measurements described above.  Representative transmittance
curves are shown in Figure~\ref{SampleTransmittance} for EJ-500 optical 
cement\footnote{The slight double absorption in the transmittance around 
550 -- 600~nm appears to be a bluish tint in the resin.  We see this to a 
greater/lesser extent in both BC-600 and EJ-500, depending on lot number.}.  
The narrow feature at 600~nm appears in all transmittance measurements 
and is assumed to be a feature of the spectrophotometer. 
We note that typical, raw measurements plateau with a 
transmittance approximately 0.80 - 0.87.  Given the refractive indexes
of the epoxies at 589~nm, we expect Fresnel losses due to reflection at the
air/material/air interfaces for ideal transmittance of 0.90--0.92.
The difference between our measurements and the ideal are attributed to
small imperfections in sample production.  Observed imperfections in the
samples fall into two broad categories and depend on sample material.  For
most samples, we observe "swirls" that appear to be differences in the
refractive index of the material which we attribute to incomplete mixing
of the resin and hardener.  Modifying mixing procedures eliminates this
effect.  The second observed imperfection is the formation of
small/microscopic bubbles in the S3GP samples during the curing process. We
observe that with smaller/thinner samples, we find smaller or no bubbles.  In 
subsequent analysis, we quote transmittance ratios for a given sample to 
remove the effects of imperfect sample production.

\begin{figure}[tb]
 \centering
 \includegraphics[width=0.45\textwidth]{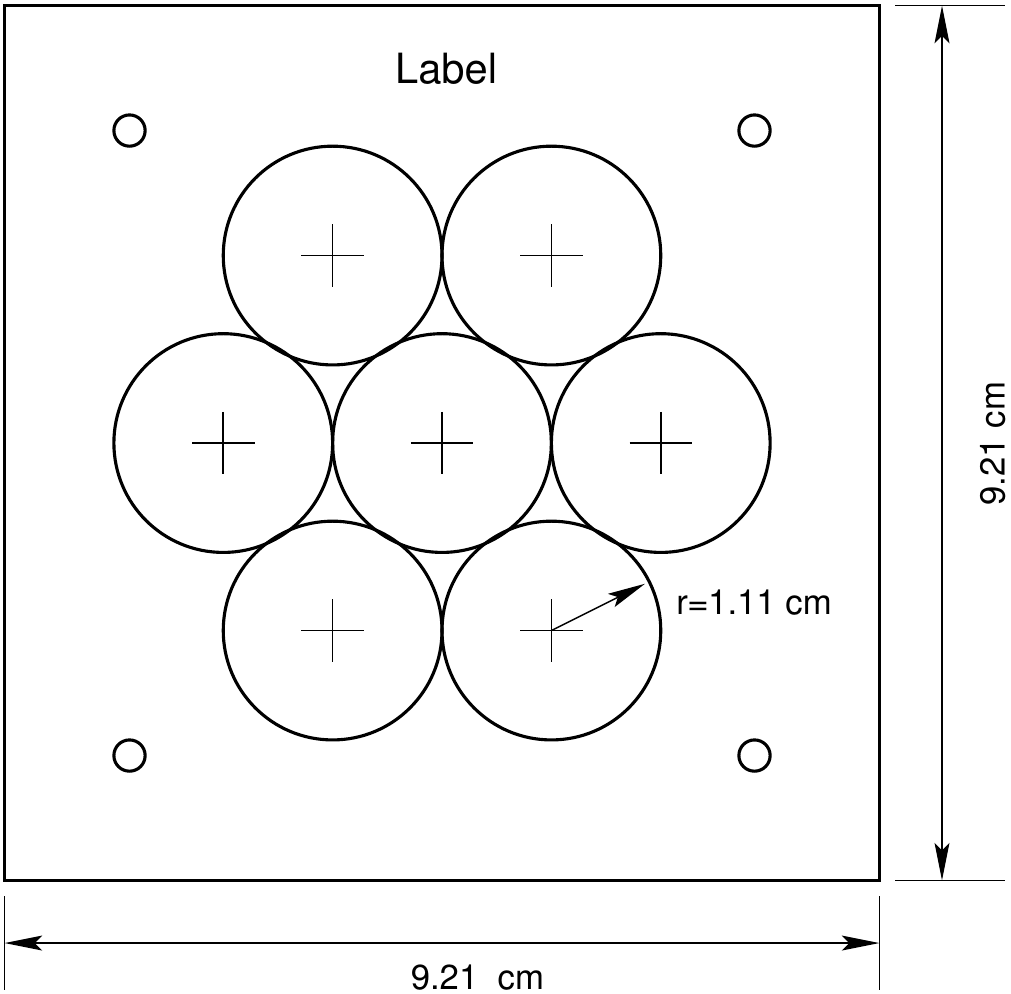}
 \hspace{0.5cm}
 \includegraphics[width=0.45\textwidth]{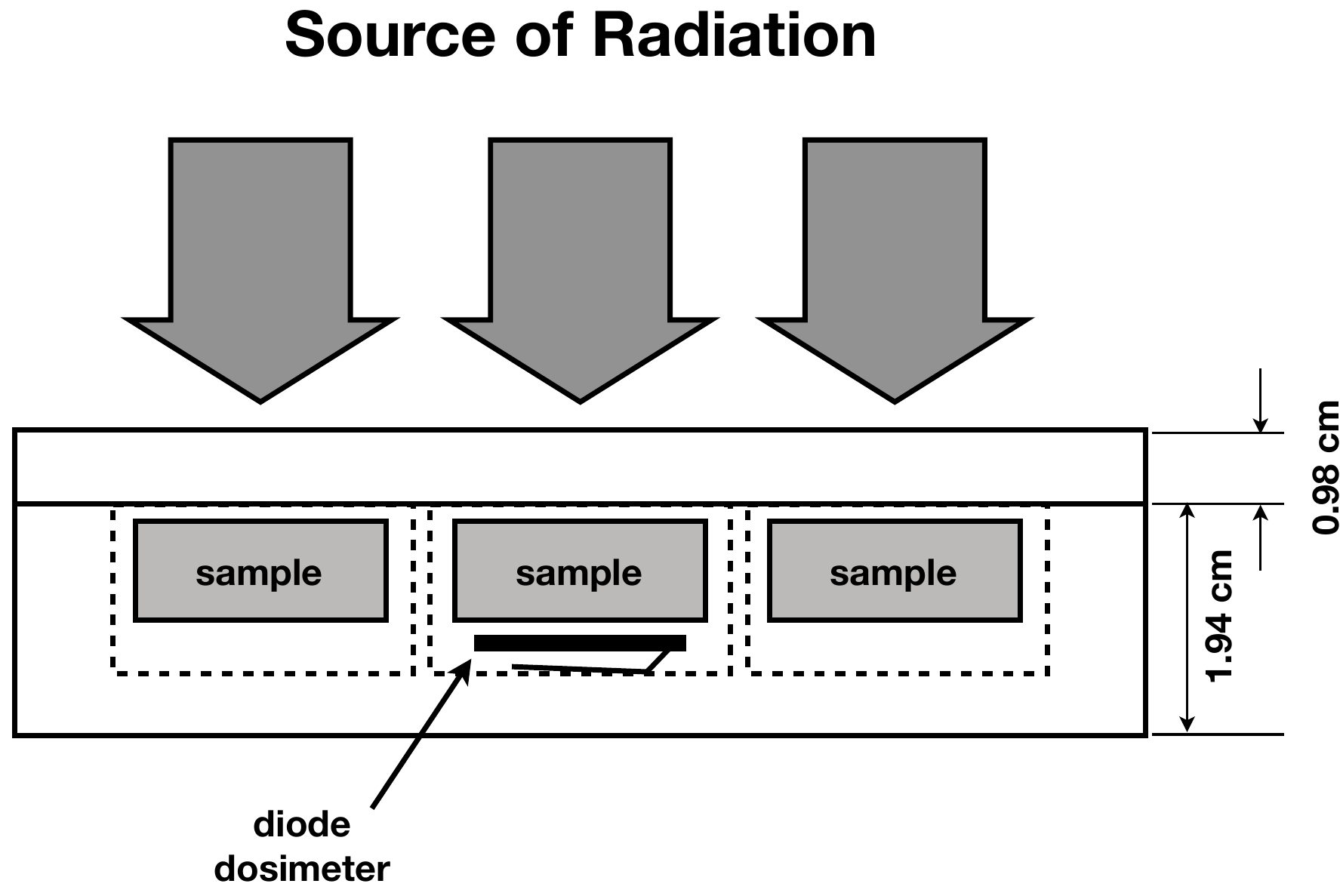}
 \caption{\label{sampleholder}Diagram of the sample holder showing the 
 seven sample wells (left) and the manner in which the cells were loaded 
 and the direction from which the samples were irradiated (right).}
\end{figure}

Once the initial transmittance spectrum was measured, a sample of each
epoxy was loaded into a seven unit, black, delrin,  sample holder.  The 
sample holders measured 9.21~cm by 9.21~cm by 1.94~cm thick.  Each
sample holder had 7, 2.22~cm diameter wells 1.17~cm deep machined in 
a hexagonal close packed arrangement with the holes spaced 1.69~cm 
on center.  A 0.98~cm thick delrin sheet lid held in placed with 4 
stainless steel screws kept the samples from moving during irradiation. 
The lid was of uniform thickness.  Figure~\ref{sampleholder} is a diagram 
of the sample holder well configuration and how the samples were loaded 
into the wells. A ${\rm 1\;cm\times 1\;cm}$ pin diode is placed under the 
center sample.  The pin diode is used as a dosimeter to measure the 
particle fluence through the sample holder.  The sample holders were then 
attached to an FR-4 strip to make a ``lolly-pop'' for installation in the
Fermilab booster for irradiation.

\section{Irradiation and Dosimetry}
\label{sec:Irradiation}
After preparation of the samples, the "lolly-pop" sample holders were installed
in the Fermilab Booster, immediately downstream of the Booster collimators 
located in the straight sections of periods 6 (collimators A and B), and 7.  
Samples were then harvested at convenient times during accelerator down times.
The radiation seen by the samples occurs from the interaction of protons 
accelerated in the booster that are far from the beam core interacting with 
the collimator material.  These interactions produce an admixture of protons, 
neutrons, photons, electrons and pions irradiating the sample.  The particle
admixture means that the irradiations may stimulate a variety of possible 
effects in the material.  This style of irradiation is ideal to survey 
radiation tolerance, but less than ideal to understand any specific effect.  

Because the samples are exposed to a mixed radiation field, we choose to 
calculate a charged particle fluence for each sample's exposure in units of 
minimum ionizing particles (MIPs) per cm$^2$. 
For the measurements reported here we use a $\rm 1\times 1~cm^2$ by 
0.02~cm thick PIN diode (Hamamatsu, model S3590-08~\cite{Hamamatsu}). 
We chose to use silicon diodes because of their observed linear increase in 
reverse bias current  with radiation exposure over a wide range of particle 
fluences.  As long as the particle admixture and spectrum remains constant 
over the exposure, this linearity is preserved and we may use a diode calibration
that yields a fluence in MIPs/cm$^2$.  The change in reverse bias 
for a silicon diode follows the relationship,
\begin{equation}
  \Phi = \frac{I_f - I_i}{\alpha_{\rm damage}V},
\end{equation}
where $I_f$, $I_i$ are the final and initial currents, $V$ is the volume of the 
diode, $\Phi$ is the charged particle fluence, the diode was exposed to 
and $\alpha_{\rm damage}$ is a damage constant.  We use the damage 
coefficient of  $3.0\times 10^{17}$~A/cm/MIP at a temperature of 20\textdegree~C
derived from radiation field measurements inside the CDF 
detector~\cite{CDFRadiationField}.  We correct all reverse bias currents 
from the temperature at which the current was measured, $T_M$, to a 
reference temperature, $T_R$, of 293.15~K (20\textdegree~C) using the relationship:
\begin{eqnarray}
  I(T_R) = \left(\frac{T_R}{T_M}\right)^2
                \exp\left\{{-\frac{1.23}{2 k_B}
                  \left(\frac{1}{T_R}-\frac{1}{T_M}\right)}\right\}\cdot I(T_M)
\end{eqnarray}
where $k_B$ is Boltzman's constant.  The factor 1.23 is the energy in eV 
between the valence and conduction bands for silicon.

The overall calibration of the diodes used a $^{137}$Cs
source to calibrate  thermal luminescent dosimeters (TLDs).  The response of the TLD 
to MIPs was then calculated using the $dE/dx$ for LiF.  The TLDs were installed 
inside the CDF detector tracking volume along with PIN 
diodes~\footnote{The original PIN diodes 
used in CDF were from the same wafers as the CDF silicon detector. These have
the same properties as those procured later from Hamamatsu.}.  The PIN diode
response was monitored as a function of exposure measured by the TLDs.  
A MARS~\cite{MARS15} simulation confirmed that the TLDs primary response was from MIPs
dominated by charged hadrons.  The overall uncertainty in this calibration
 was found to be approximately 5\% including all systematics.  Details of
this process may be found elsewhere~\cite{CDFRadiationField}.

\begin{figure}[tb]
 \centering
 \includegraphics[width=\textwidth]{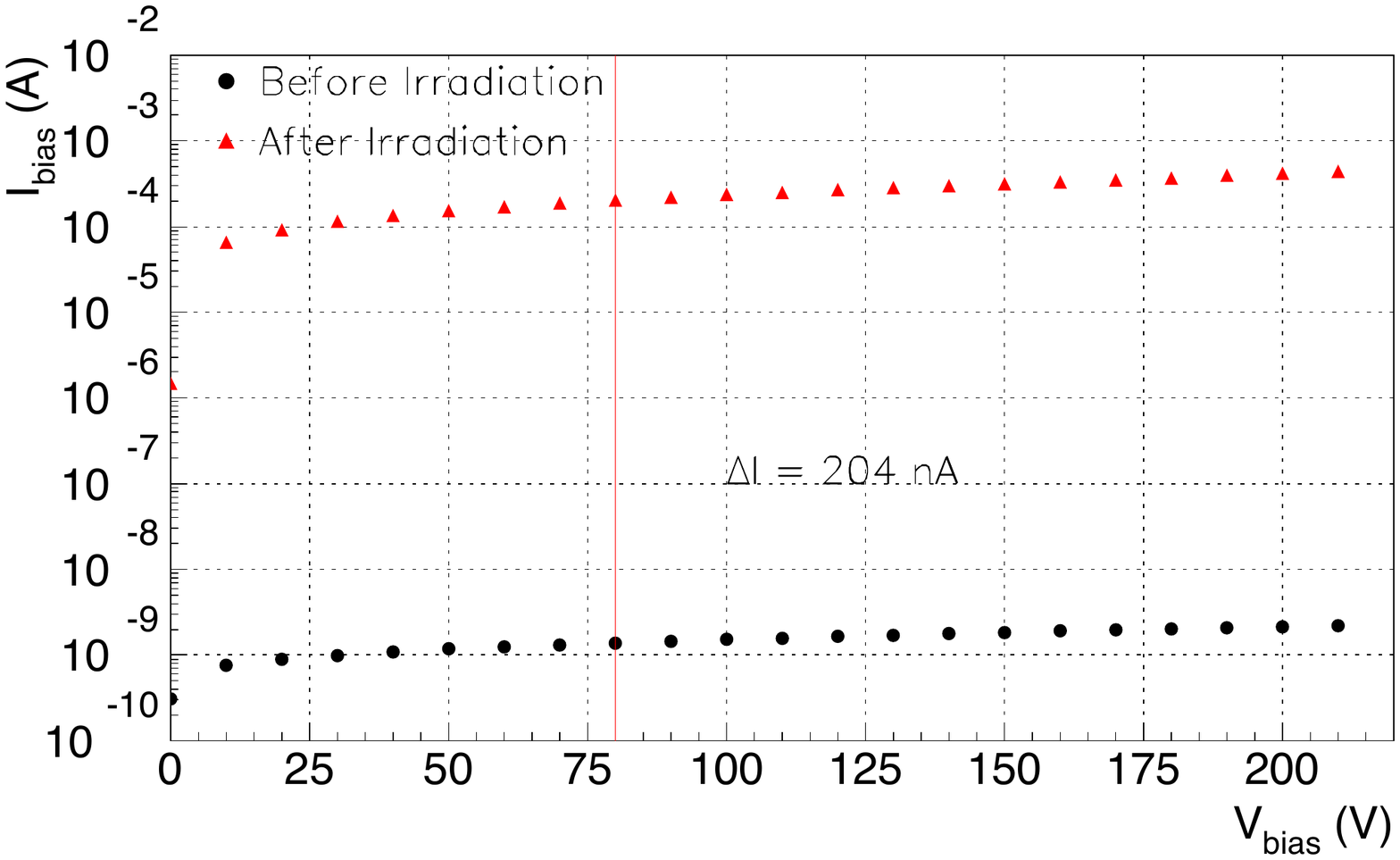}
 \caption{\label{IVcurve}Raw reverse bias current vs voltage data for a 
  single diode dosimeter before (circles) and after (triangles) irradiation.  
  The change in current is shown for 80~V as indicated by the vertical line.  
  After temperature correcting the two measurements the current 
  change is reduced to $174\;\mu$A. The corresponding 
  fluence from this irradiation is $2.9\times 10^{14}$~MIPs/cm$^2$.}
\end{figure}

The reverse bias current for a given diode is obtained by measuring its I-V 
curve using a program running on a laptop that controls 
a Keithley model 6517A electrometer~\cite{keithley6517a}.  The
program set the bias voltage (V) and allowed the current (I) to settle for 
5~seconds before recording both the applied voltage and current.  The 
process was repeated in 10~V intervals over the range 0 -- 210~V.  At the
time of the I-V measurements both the ambient temperature and humidity
were recorded for use later.  Ambient temperatures during the measurements
varied over the range (18.9\textdegree -- 27.7\textdegree~C).  Typical 
raw I-V curves are shown in 
Figure~\ref{IVcurve} for both before and after irradiation.  From the 
measurements we calculate the change in current, after - before irradiation, 
for a bias voltage of 80~V for the dosimetry measurements.  For the conversion
from fluence (MIPs/cm$^2$) to absorbed dose (Gy), we use the value of $dE/dx$ 
for acrylic as this polymer most closely matches the density of the epoxy samples.

%

\section{Transmittance Analysis}
\label{sec:analysis}
\begin{figure}[htbp]
 \centering
  \includegraphics[width=0.45\textwidth]{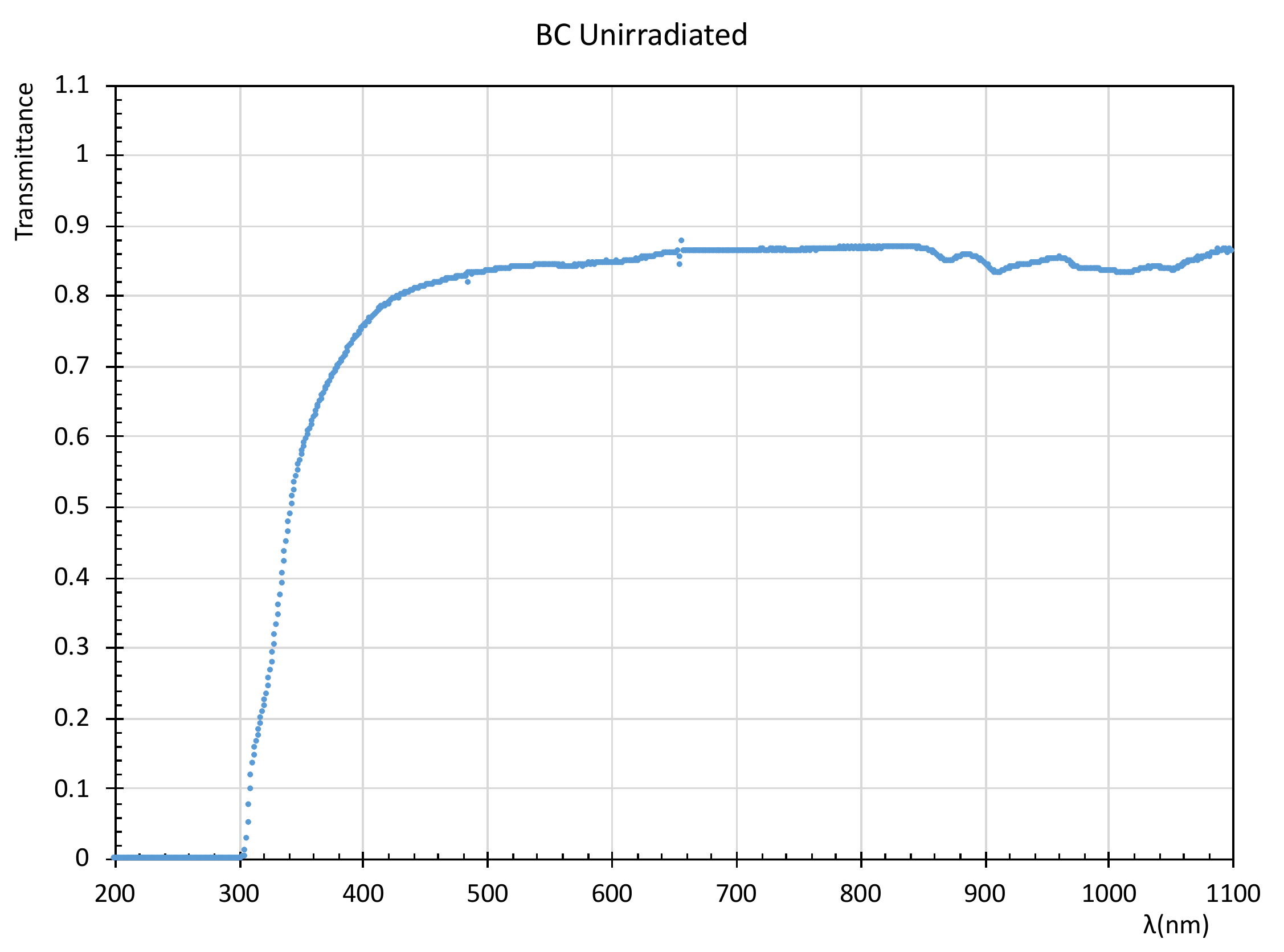}
  \includegraphics[width=0.45\textwidth]{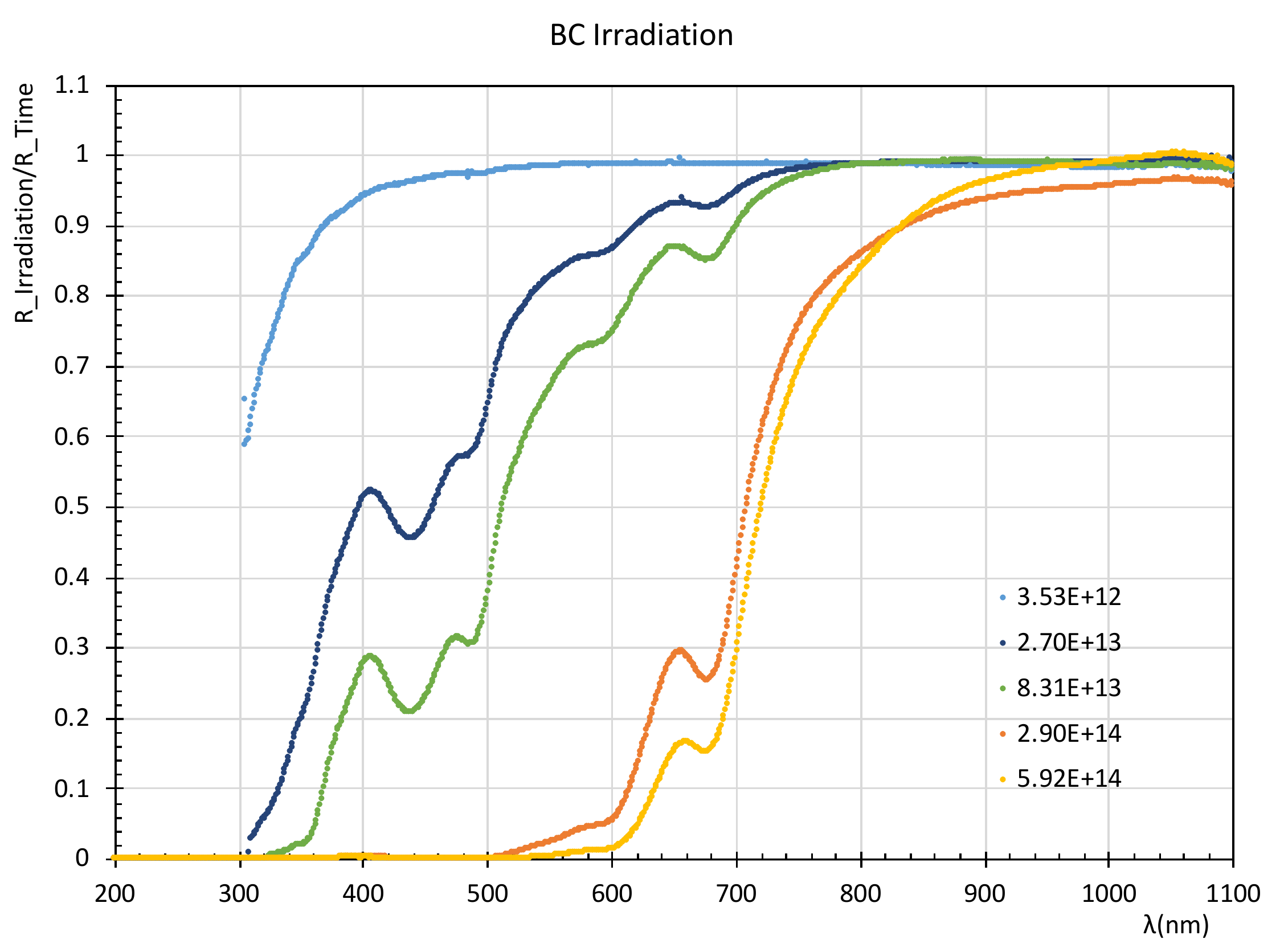} \\
  \includegraphics[width=0.45\textwidth]{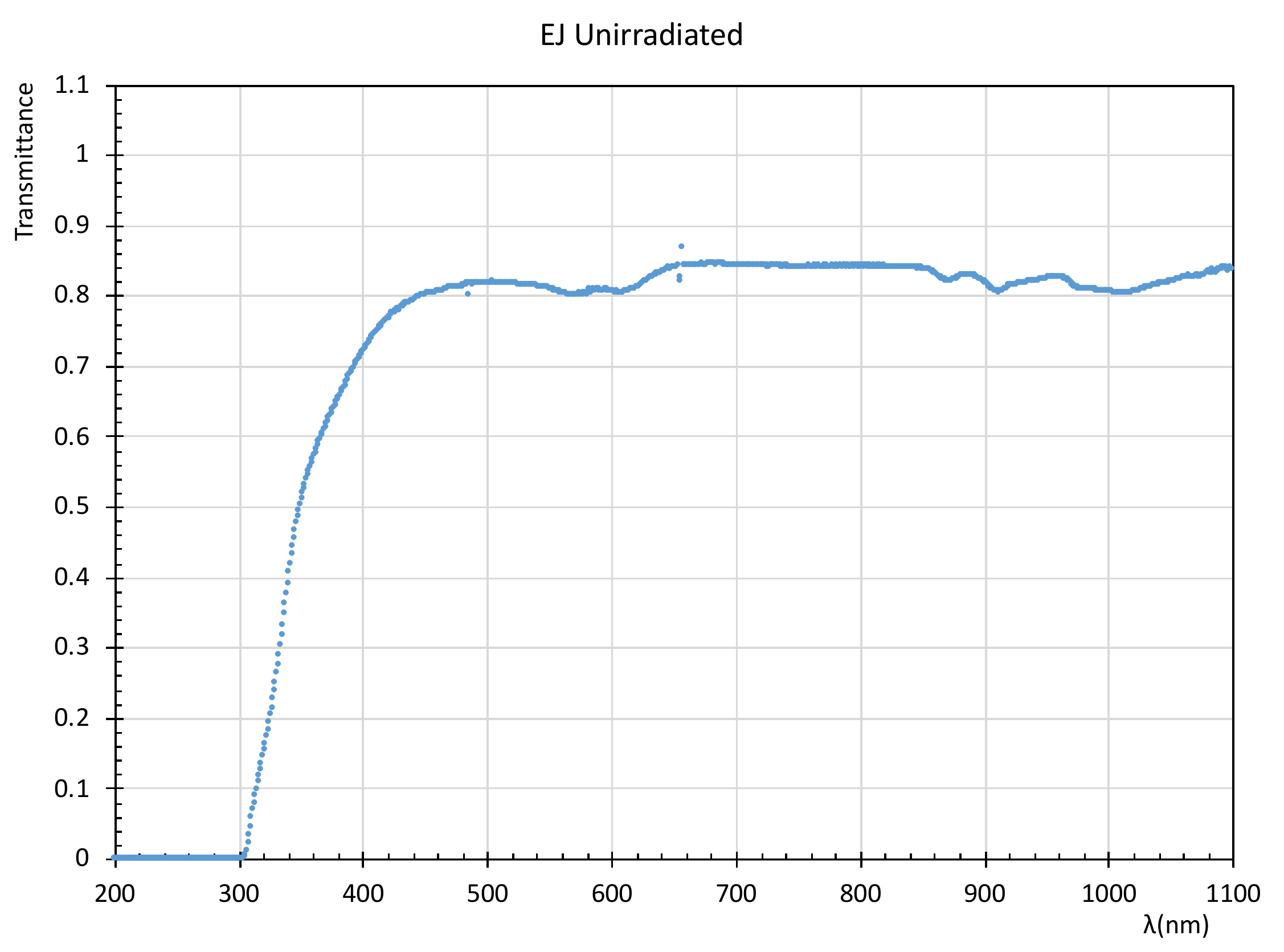}
  \includegraphics[width=0.45\textwidth]{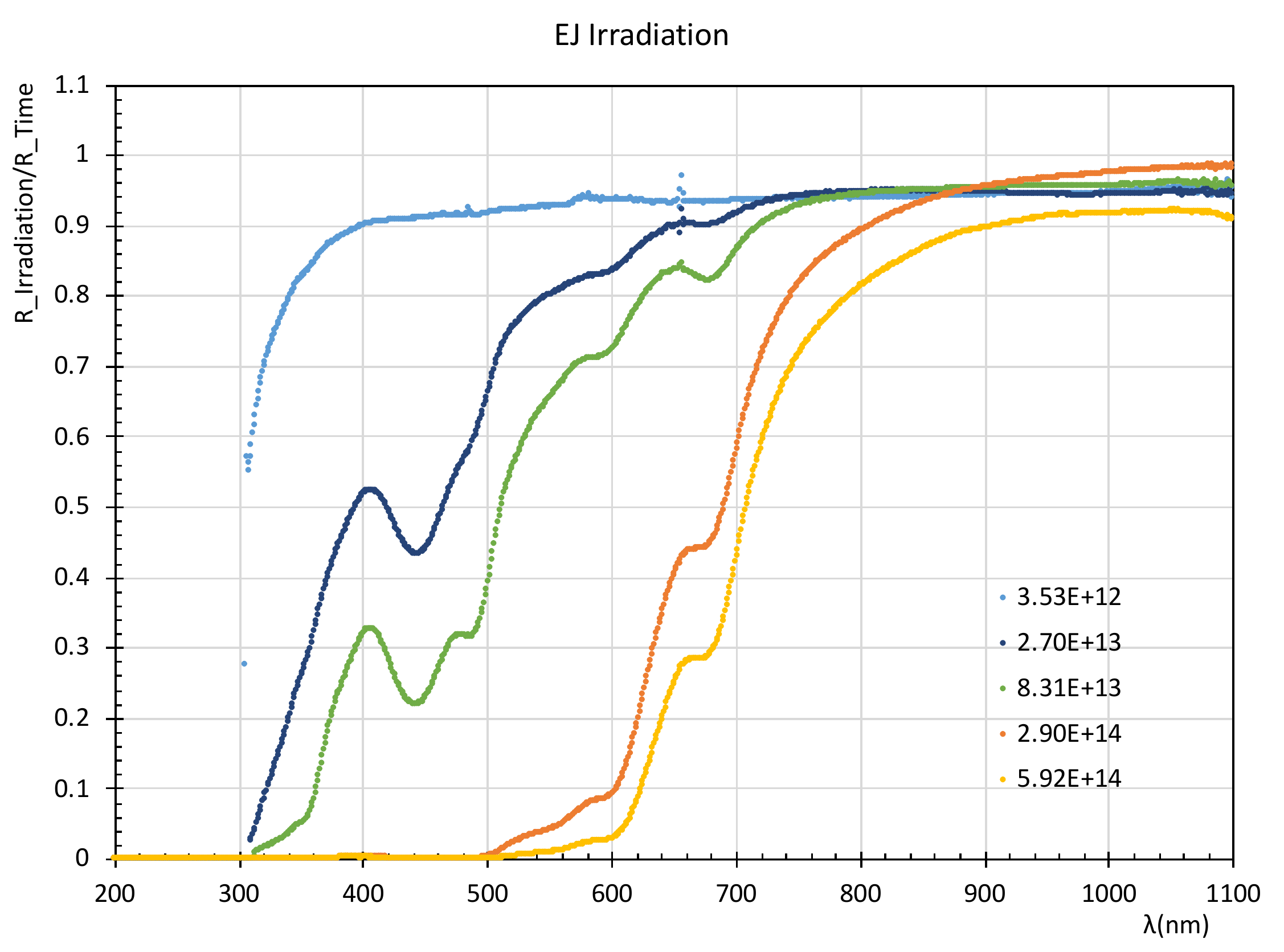} \\
  \includegraphics[width=0.45\textwidth]{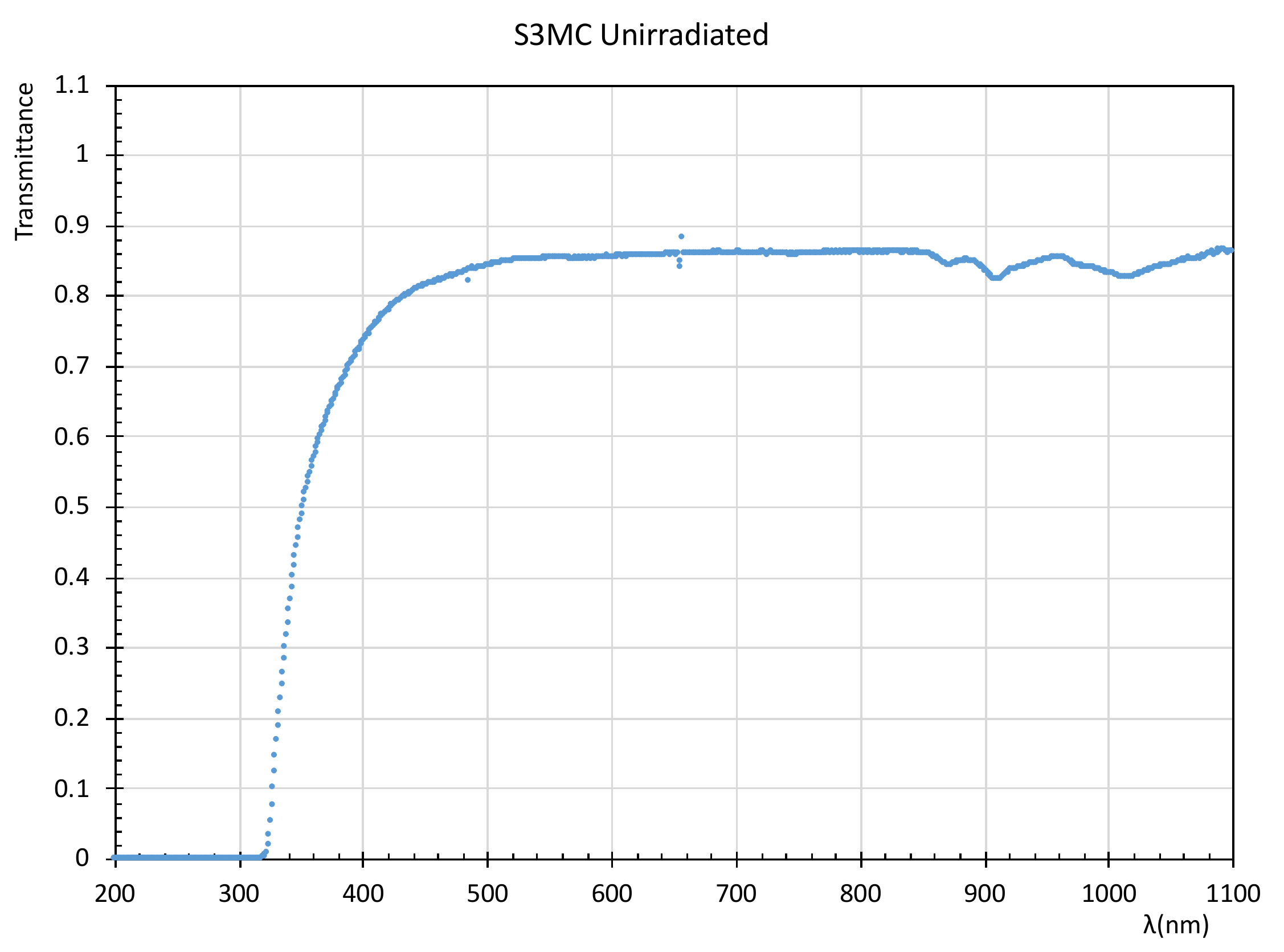}
  \includegraphics[width=0.45\textwidth]{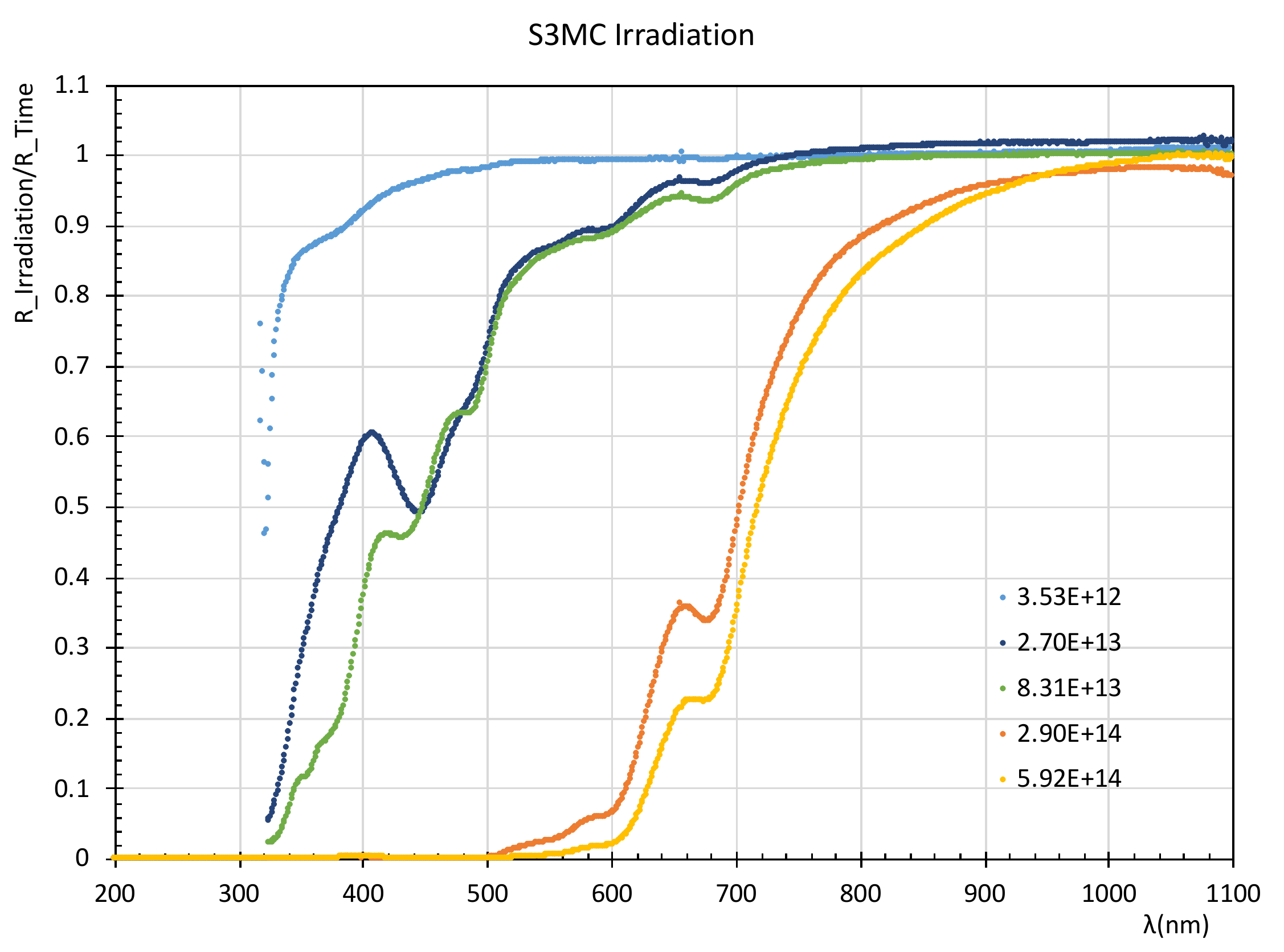} \\
  \includegraphics[width=0.45\textwidth]{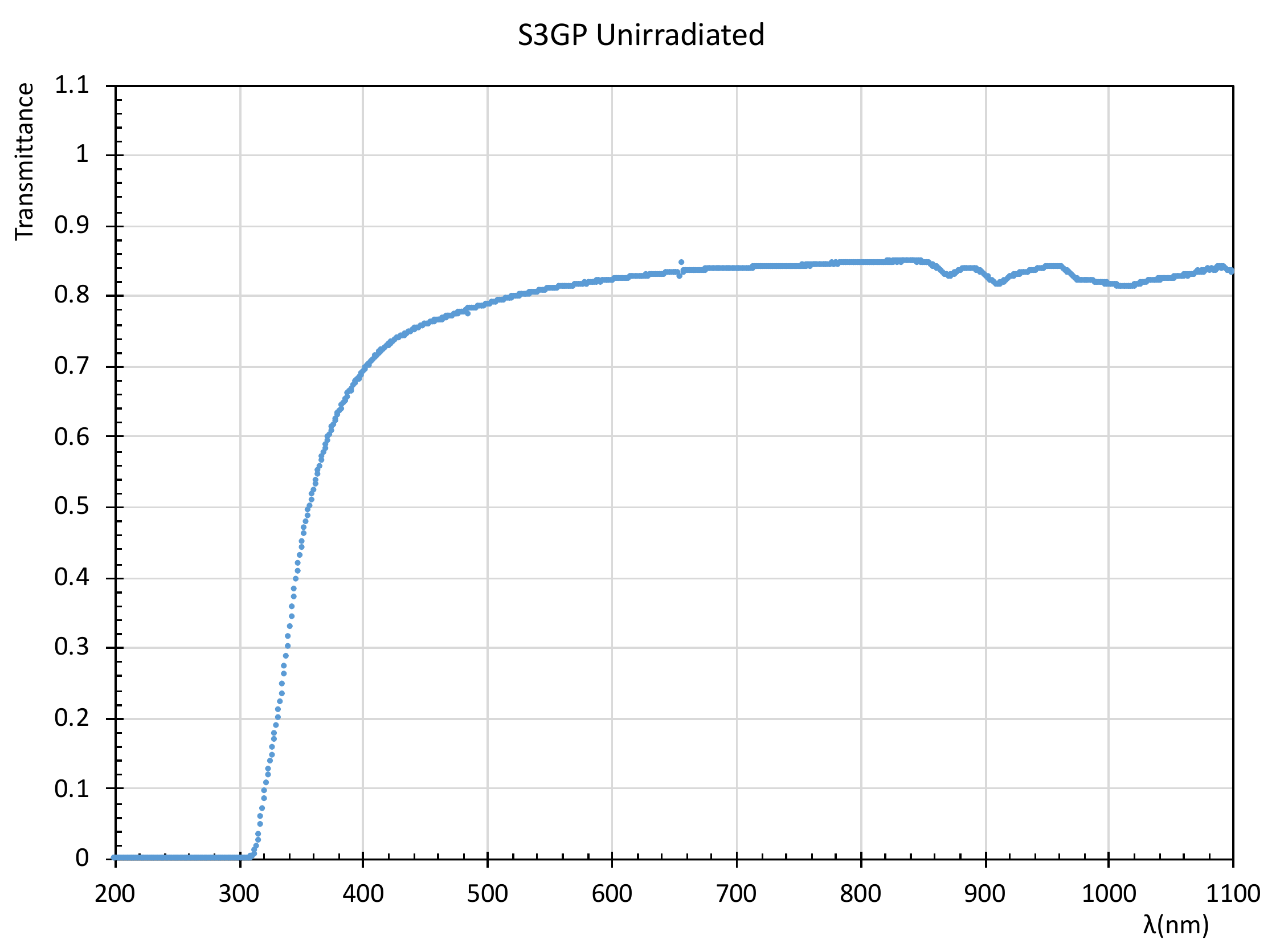} 
  \includegraphics[width=0.45\textwidth]{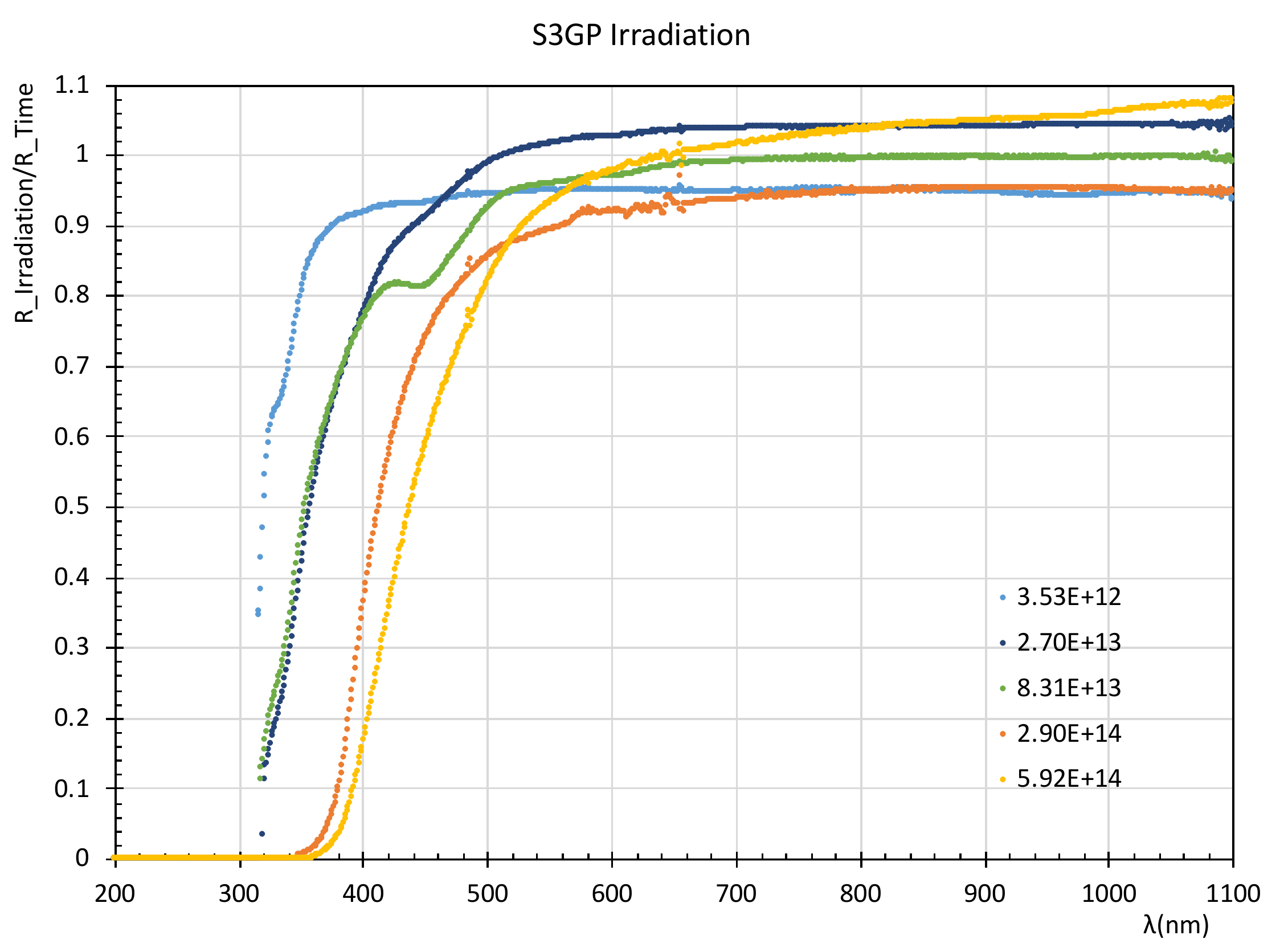} \\
  \caption{\label{SampleSpectra1}Transmittance spectrum (left column) and 
  effect of radiation (right column) for four samples of optical cements.
  Fractional uncertainties of 5\% (right column) and 10\% (left column) for each 
  point were omitted for clarity.}
\end{figure}

\begin{figure}[htb]
 \centering
  \includegraphics[width=0.45\textwidth]{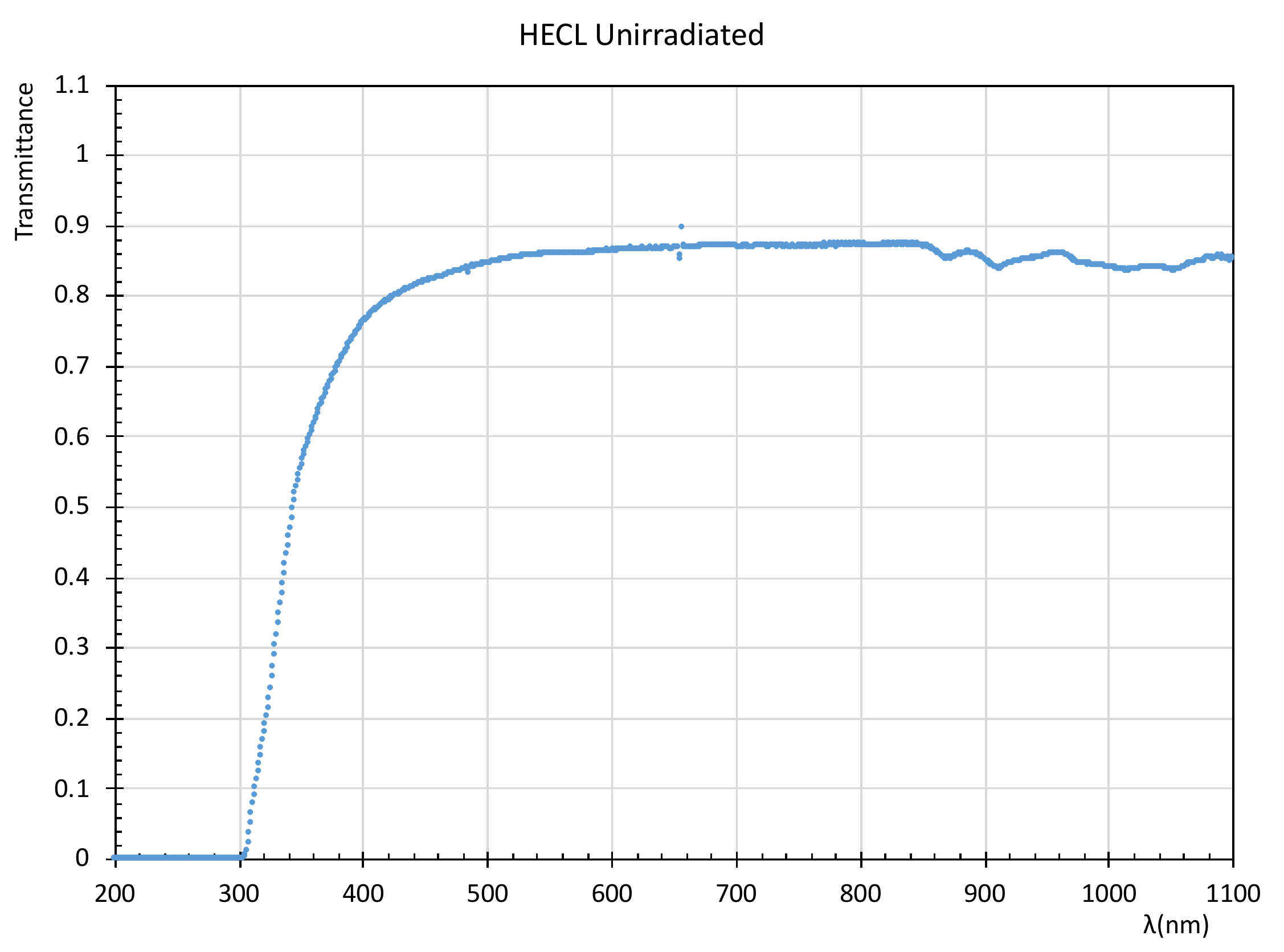}
  \includegraphics[width=0.45\textwidth]{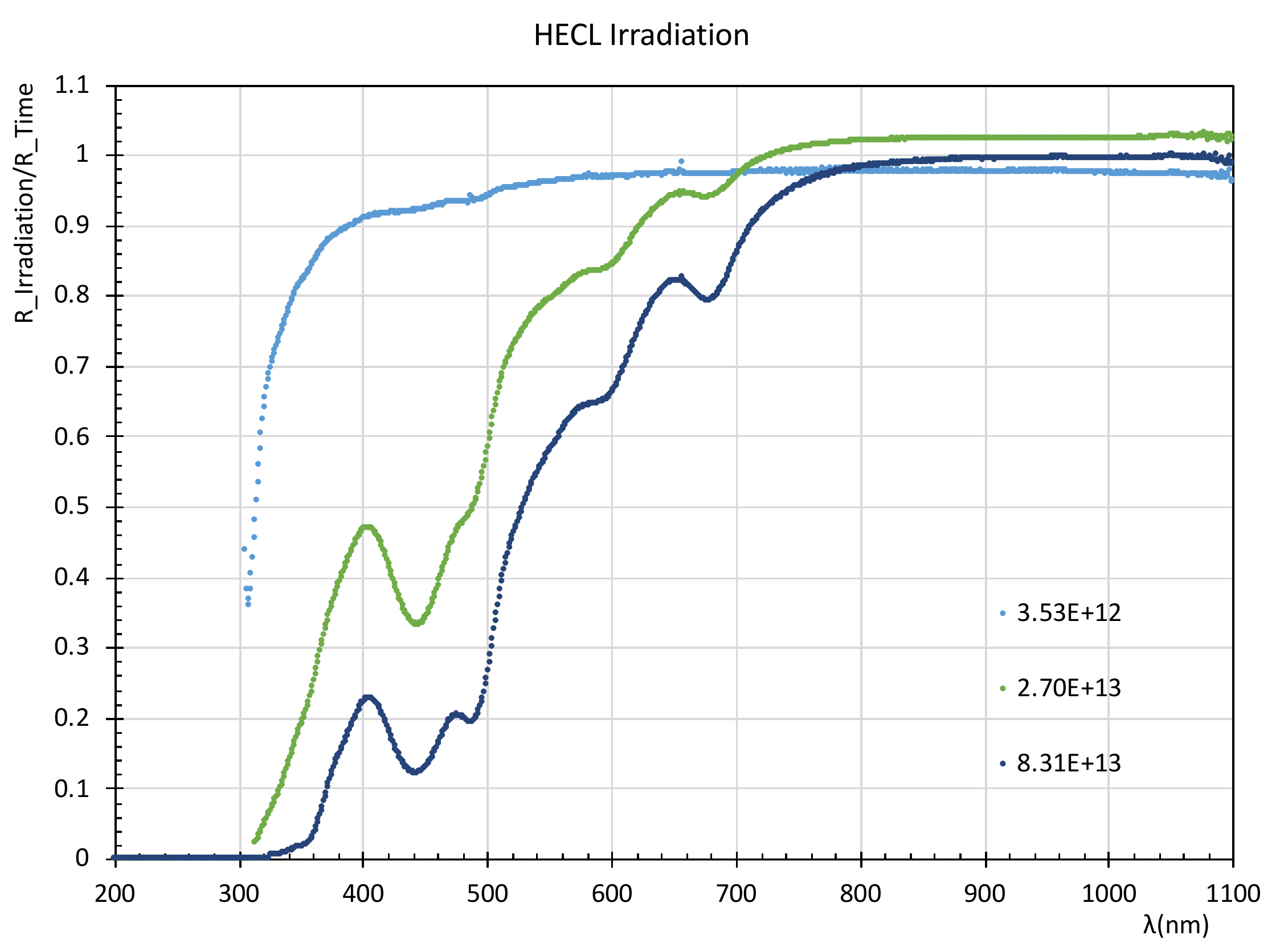} \\
  \includegraphics[width=0.45\textwidth]{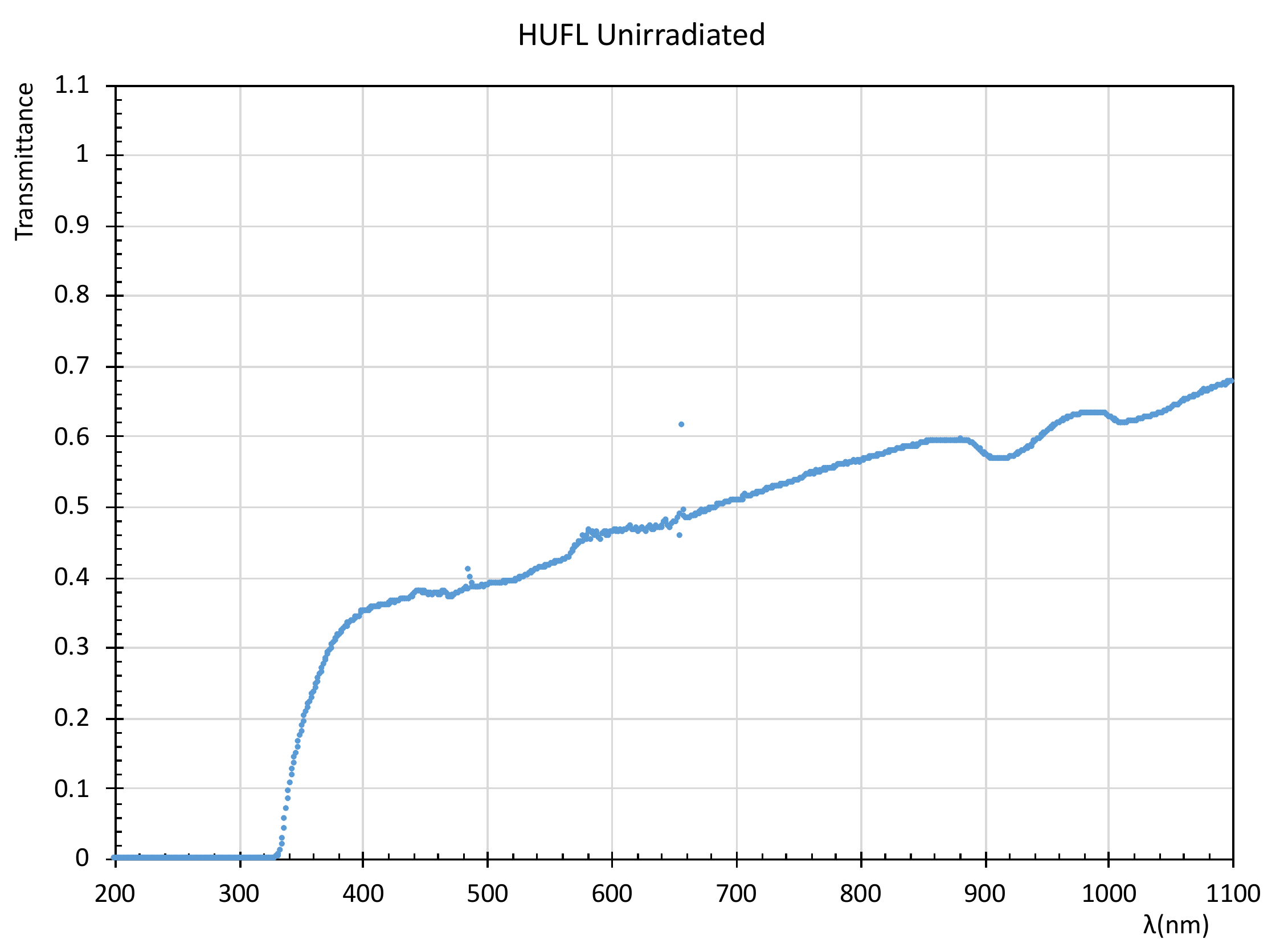}
  \includegraphics[width=0.45\textwidth]{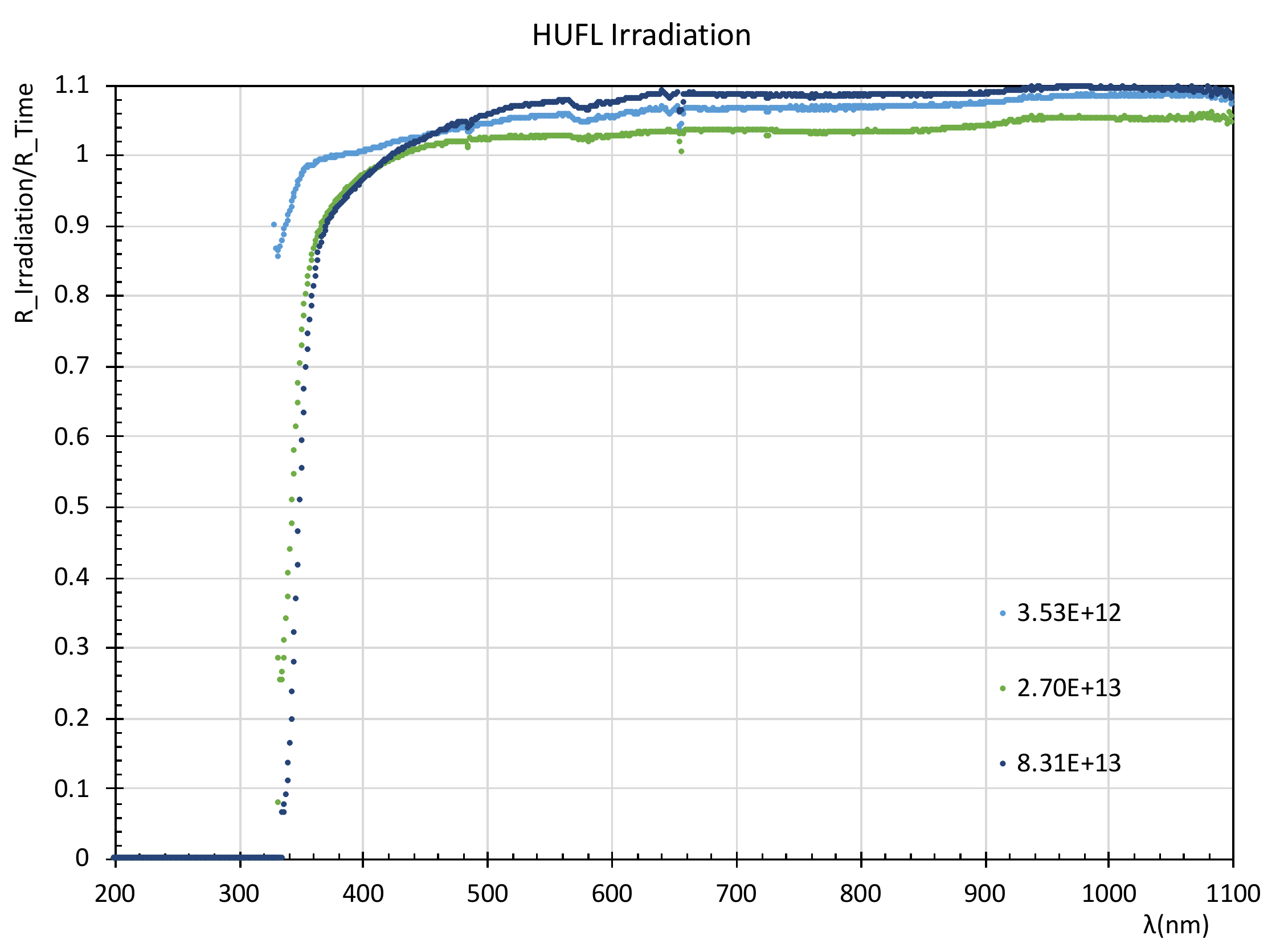} \\
  \caption{\label{SampleSpectra2}Transmittance spectrum (left column) and
   effect of radiation (right column) for HECL and HUFL samples of optical cements. 
   Fractional uncertainties of 5\% (right column) and 10\% (left column) for each 
   point were omitted for clarity.}
\end{figure}

At completion of each irradiation, the sample holders were harvested, 
dosimetry performed for the specific sample holder and the transmittance was 
measured for both the irradiated in the holder and the control samples 
that were unirradiated.  We are interested in changes in the transmittance 
between when an adhesive sample was made and after its irradiation.  
Calculation of the ratio of the transmittance after irradiation to that measured 
before irradiation allows us to quantify the change in transmittance observed.
However, the change in transmittance may include effects due to
radiation and aging of the material.  We therefore calculate the double
ratio of transmittances for irradiated and unirradiated control samples 
using the equation:
\begin{equation}
  R^i_{d,t}(\lambda) = \frac{T^i_{d,t}(\lambda)/T^i_{0,0}(\lambda)}{
                       T^r_{0,t}(\lambda)/T^r_{0,0}(\lambda)}
\end{equation}
where the $T^i_{d,t}(\lambda)$ are measured transmittances for each sample, 
$i$, at a given absorbed dose, $d$, and time, $t$.  The ``0'' dose correspond
to control samples that were kept in a cabinet.  The control samples saw
background radiation only (mostly cosmic rays).  The ratio in the denominator
is the relative transmittance due to aging.  The ratio in the numerator is
the relative transmittance due to aging and irradiation. 
Figures~\ref{SampleSpectra1} and~\ref{SampleSpectra2} show the 
initial transmittance of a sample (left column) and the transmittance double 
ratio (right column) for the six samples studied.  Uncertainties are omitted
from the figures for clarity.  Typical uncertainties for the transmittance and
double ratio at 1100~nm are $\pm$~5\% and $\pm$~10\%, respectively.
The effects of radiation show up as a loss of transmittance at 
short wavelengths for all samples.  Higher doses leading to a loss of 
transmittance at longer wavelengths.  Many of the spectra also show 
structure with two broad absorption dips in the region between 
450--500~nm and another at approximately 680~nm.  Two samples,
S3GP and the HUFL, stand out showing less effect due to radiation.
The careful reader will note that in some samples the transmittance
double ratio, $R^i_{d,t}$ curve,  exceeds unity at the longer wavelengths.
For all samples the curves lie within the 10\% uncertainty envelope 
quoted above.  This uncertainty envelope is dominated by variations 
within each sample (ref, Figure~\ref{SampleTransmittance}).  Because 
the non-uniformity is due to small, local defects this leads to a larger 
variation in the reproducibility for that sample. No clear pattern in 
the transmittance double ratio is seen with radiation exposure for any 
adhesive sample.  For the S3GP and HUFL epoxies, the control sample 
had poorer transmittance and uniformity than seen in the other epoxies.

\begin{figure}[htb]
 \centering
 \includegraphics[width=0.49\textwidth]{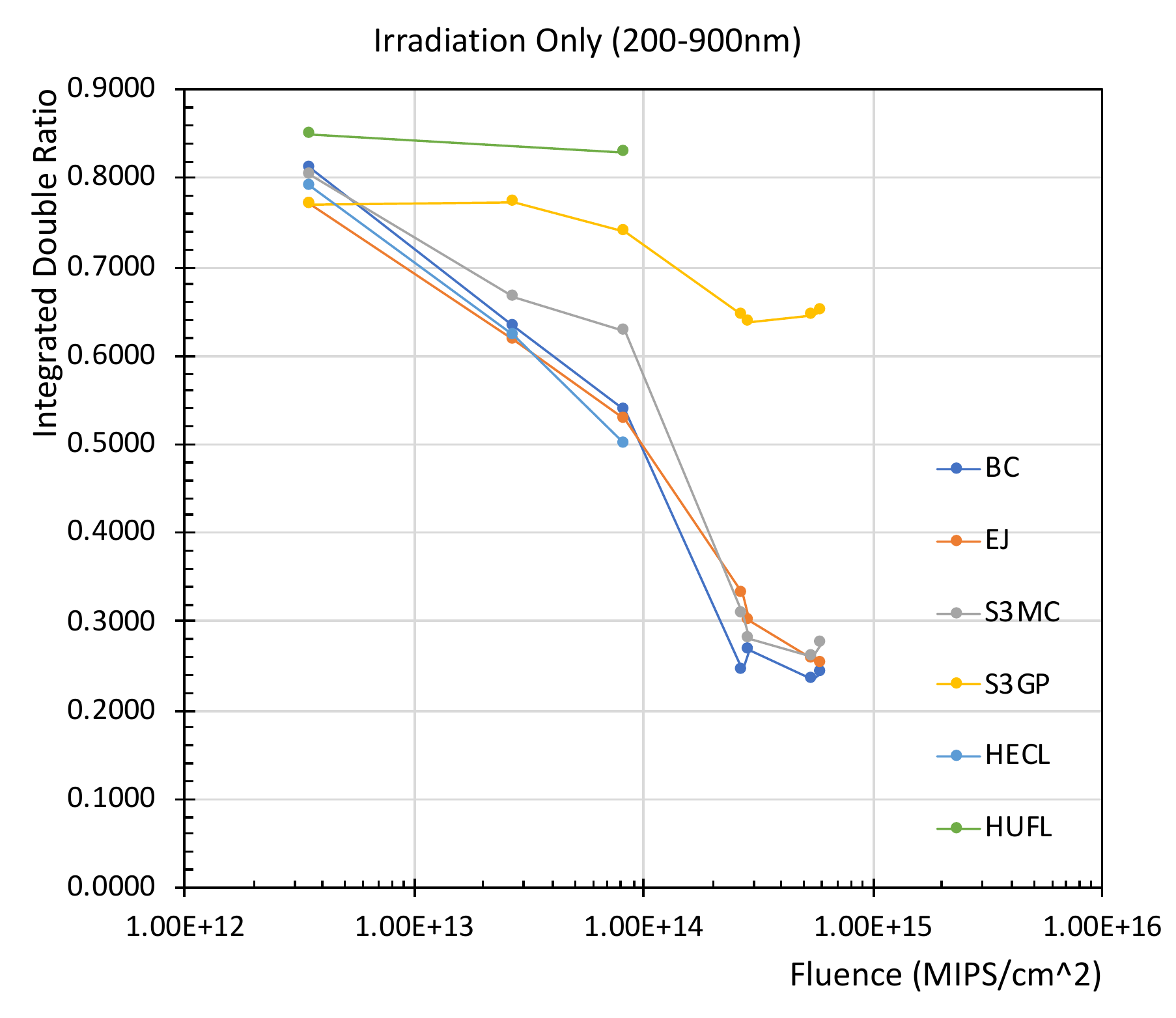}
 \includegraphics[width=0.49\textwidth]{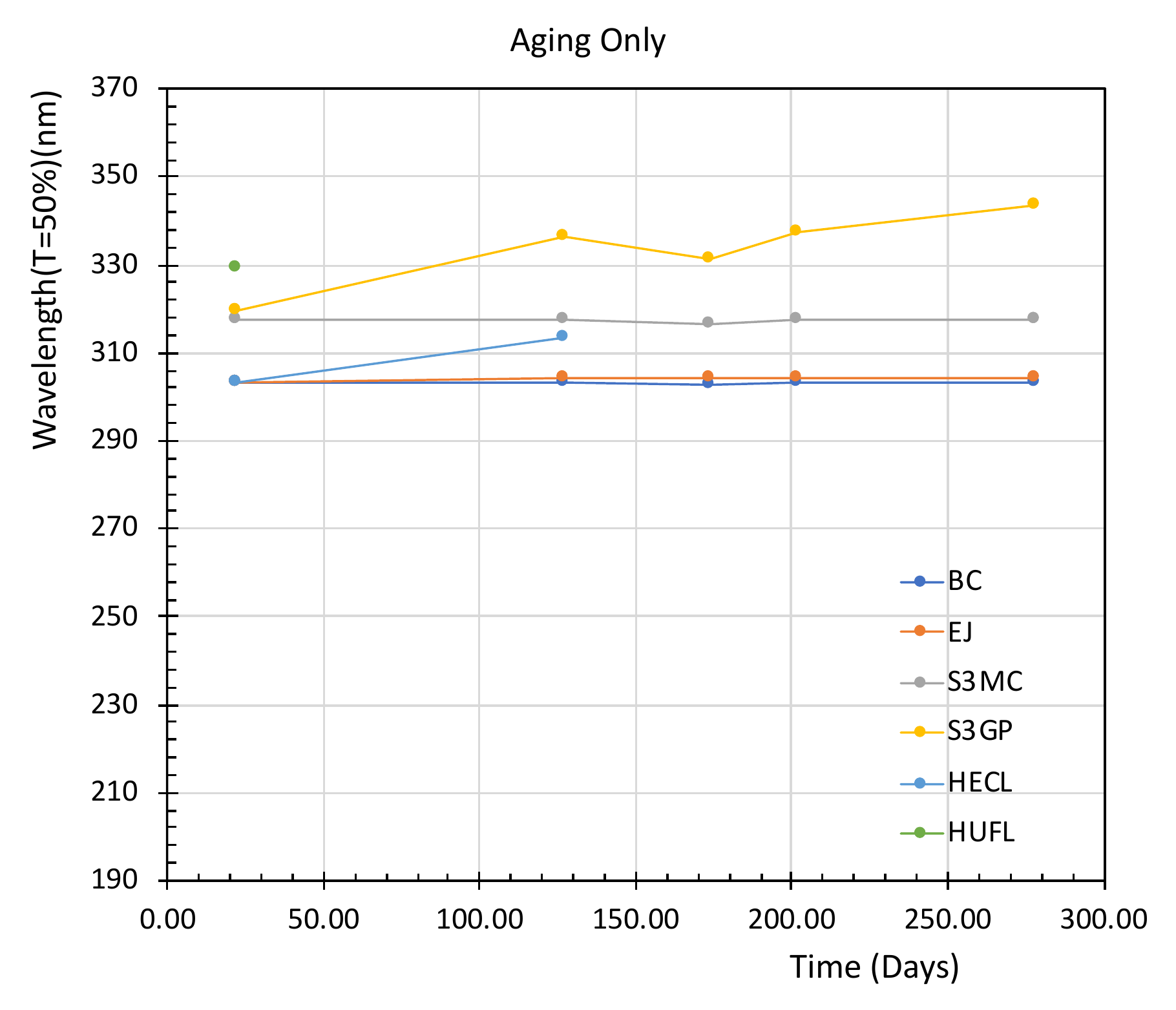}
 \caption{\label{sampledoubleratio}Results of Irradiation study (left) 
  and aging (right) studies. Uncertainties in the two figures are the size 
  of the points.}
\end{figure}
A figure of merit is needed to more compactly and directly compare 
the samples against one another.  From the features observed, we
choose to integrate the measured double ratio curve over the wavelength
range 200--900~nm.  The integrals are all normalized over the same
wavelength range.  This figure of merit represents the expected 
transmittance to a detector with flat sensitivity over this wavelength 
range.  One may similarly integrate the ratio, 
$T^r_{0,t}(\lambda)/T^r_{0,0}(\lambda)$,
to extract a similar effect in the aging of the control samples.  Due to the
non-uniformity of the control sample, the integral hid differences in 
sample coloration visible by eye.
Consequently, to illustrate the effects of aging we choose to quote the 
50\% threshold wavelength as a function of time.
Figure~\ref{sampledoubleratio} shows the double ratio integral as a 
function of the radiation exposure (left) and the 
transmittance ratio 50\% threshold wavelength as a function of time for 
the control samples (right).  The Figure shows 
considerably better transmittance for the two adhesives noted above.
Note that the S3GP sample also shows an increase of the threshold 
wavelength with sample age.

Examining the samples significantly damaged during irradiation, we 
observe a clearing of those sample at their edges.  This clearing is 
parallel to all sample surfaces with the clearing front extending a fraction 
of a mm after several months.
This clearing is also observed in thin films of  epoxy from detectors when 
a joint is broken and the epoxy is exposed to the ambient environment.
The clearing of the thin film of epoxy occurs over a few days.  
This clearing effect is similar to that seen in plastic scintillator when 
exposed to air or oxygen after irradiation~\cite{Bross1992}.  At this 
time we have not isolated the agent responsible for the effect.

\section{Discussion}
\label{sec:discussion}
The samples prepared in this study are approximately 7~mm thick which 
is considerably thicker than a typical adhesive joint.  
We found that for most scintillator detectors, a typical joint thickness 
was  0.1~mm.  Under the assumption that light absorption in the material 
follows a single exponential decay, for the 7~mm thick samples the 
amount of light lost at the 50\% threshold wavelength corresponds to 
about 0.4\% for a 0.1~mm thick adhesive joint.  A 0.4\% loss of light
is insignificant for most measurements. Figures~\ref{SampleSpectra1}
and~\ref{SampleSpectra2} show regions where the transmittance or
double ratio is indistinguishable from zero.  The Agilent spectrophotometer
used for this analysis has a readout threshold for transmittance of 
$2\times 10^{-4}$.  The threshold transmittance for the sample 
thicknesses approximately 7~mm thick would mean a 5\% loss of 
light at that wavelength for the 0.1~mm thick adhesive joint.  For 
many of the epoxy samples studied at the highest fluences, we find the 
transmittance below the readout threshold for wavelengths shorter
than 400~nm indicating at least 5\% light loss at those wavelengths.
This means a substantial loss of light in the wavelength regions where 
typical photomultiplier tubes used in scintillator detectors have their 
peak quantum efficiency.

\section{Summary}
\label{sec:summary}
We have measured the radiation tolerance of a variety of commercially
available optical epoxies that may be used to couple plastic scintillator 
to a light sensor as a function of radiation exposure.  Our studies
recorded the transmittance spectrum as a function of charged particle
fluence measured in the mixed radiation environment near the Fermilab
booster collimators.  Most of the epoxies studied show reduced light
transmission at short wavelengths for a given radiation exposure.  When
the radiation exposure is increased, transmittance at successively longer 
wavelengths is attenuated. For exposure fluences above 
$3\times 10^{14}$~MIPs/cm$^2$, 
our transmittance measurements are less than the detection threshold for 
wavelengths below 500~nm for three of the six samples tested.  Extrapolating 
from our 7~mm thick samples to a glue thickness of 0.1~mm means 
at least a 5\% light loss in a real detector in the same wavelength range.
However, for two of our samples, we find considerably better light 
transmission at wavelengths near 400~nm.  One of these epoxies proved
difficult to work with, HUFL. The other epoxy, S3GP, while showing signs 
of yellowing with age, showed little degradation for fluences up to 
$5.9\times10^{14}$~MIPs/cm$^2$.  Our studies indicate that S3GP promises
to be a good candidate for bonding scintillator to light guides and
photosensors in the future, though additional studies are needed to
quantify the effects of aging with this epoxy.


\acknowledgments
Work supported by the Fermi National Accelerator Laboratory, managed and
operated by Fermi Research Alliance, LLC under Contract No. DE-AC02-07CH11359 with the U.S. Department of Energy,  The U.S. Government 
retains the publisher, by accepting this article for publication,
acknowledges that the United States Government retains a non-exclusive, 
paid-up, irrevocable, world-wide license to publish or reproduce the
published form of this manuscript, or allow others to do so, for United States Government purposes.


\end{document}